\documentclass{cupconf}
\usepackage{graphicx}

  \checkfont{eurm10}
  \iffontfound
    \IfFileExists{upmath.sty}
      {\typeout{^^JFound AMS Euler Roman fonts on the system,
                   using the 'upmath' package.^^J}%
       \usepackage{upmath}}
      {\typeout{^^JFound AMS Euler Roman fonts on the system, but you
                   dont seem to have the}%
       \typeout{'upmath' package installed. cupconf.cls can take advantage
                 of these fonts,^^Jif you use 'upmath' package.^^J}%
      }
  \else
  \fi

  \checkfont{msam10}
  \iffontfound
    \IfFileExists{amssymb.sty}
      {\typeout{^^JFound AMS Symbol fonts on the system, using the
                'amssymb' package.^^J}%
       \usepackage{amssymb}%
         \let\leq=\leqslant
         
      }{}
  \fi

  \IfFileExists{amsbsy.sty}
    {\typeout{^^JFound the 'amsbsy' package on the system, using it.^^J}%
     \usepackage{amsbsy}}
    {}

\newsavebox{\astrutbox}
\sbox{\astrutbox}{\rule[-5pt]{0pt}{20pt}}

\title[Reverberation Mapping of AGNs]{Reverberation mapping of
active galactic nuclei}

\author[B.M. Peterson and K. Horne]%
{B.\ns M.\ns P\ls E\ls T\ls E\ls R\ls S\ls O\ls N$^1$
\and K.\ns H\ls O\ls R\ls N\ls E$^2$}

\affiliation{$^1$Department of Astronomy, The Ohio State University,
140 West 18th Avenue, Columbus, OH, USA\\[\affilskip]
$^2$School of Physics and Astronomy, University of St.\ Andrews,
St.\ Andrews KY16 9SS, Scotland}

\pubyear{2004}
\volume{538}
\pagerange{1--10}
\date{?? and in revised form ??}
\begin{document}

\maketitle

\begin{abstract}
Reverberation mapping is a proven technique that is used
to measure the size of the broad emission-line region 
and central black hole mass in active galactic nuclei.
More ambitious reverberation mapping programs that are 
well within the capabilities of {\em Hubble Space Telescope}
could allow us to determine the nature and
flow of line-emitting gas in active nuclei and 
to assess accurately 
the systematic uncertainties in reverberation-based
black hole mass measurements.
\end{abstract}

\firstsection 

\section {Introduction: The Inner Structure of AGNs}

There is now general consensus that the long-standing paradigm for
active galactic nuclei (AGNs) is basically correct, i.e., that AGNs
are fundamentally powered by gravitational accretion onto supermassive
collapsed objects. Details of the inner structure of AGNs, however,
remain sketchy, although both emission lines and absorption lines
reveal the presence of large-scale gas flows on scales of
hundreds to thousands of gravitational radii. The
accretion disk produces a time-variable high-energy
continuum that ionizes and heats this nuclear gas, and the broad
emission-line fluxes respond to the changes in the illuminating flux from
the continuum source.  The geometry and kinematics of the broad-line
region (BLR), and fundamentally its role in the accretion process, are
not understood. Immediate prospects for understanding this key element
of AGN structure do not seem especially promising with the realization
that the angular size of the nuclear regions projects to only
microarcsecond scales even in the case of the nearest AGNs.
Unfortunately, there is only very limited information about the BLR
from the emission-line profiles alone, since many simple kinematic
models are highly degenerate.  Nevertheless, it has been possible to
draw a few basic interferences about the nature of the BLR:
\begin{enumerate}
\item {\it There is strong evidence for a disk component in at least
some AGNs.} In particular, there is a relatively small subset of AGNs
whose spectra show double-peaked Balmer-line profiles. Double-peaked
profiles are generally associated with rotating Keplerian disks.
\item {\it There is strong evidence for an outflowing component in
many AGNs.} Some emission lines have strong blueward asymmetries,
suggesting that we preferentially observe outflowing material on the
nearer side of an AGN. Slightly blueshifted (relative to the
systemic redshift of the host galaxy) absorption features are quite
common in AGNs, and there is a good deal of evidence that this
absorption, seen primarily in ultraviolet and X-ray spectra,
arises on scales similar to that of the broad emission lines.
\item{\it There is strong evidence that gravitational acceleration by
the central source is important.} As discussed below, a physical scale
for the size of the line-emitting region can be obtained by the
process of reverberation mapping. The derived scale length for each
line is different, with lines that are characteristic of
high-ionization gas arising closer to the central source than lines
that are more characteristic of low-ionization gas, 
thus demonstrating ionization
stratification within the BLR.  Moreover, the higher ionization lines are
broader, and indeed the relationship between size and velocity
dispersion of the line-emitting region shows a virial-like
relationship, i.e., $r \propto \Delta V^{-2}$, where $r$ is the
characteristic scale for a line which has Doppler width $\Delta V$.
\end{enumerate}

The conclusion that gravity is important leads us directly to
an estimate of the black hole mass, which we take to be
\begin{equation}
M_{\rm BH} = \frac{f r \Delta V^{2}}{G},
\end{equation}
where $G$ is the gravitational constant and $f$ is a scaling factor of
order unity that depends on the presently unknown geometry and
kinematics of the BLR.

In this brief introduction, we already see the two major reasons that
understanding the BLR is of critical importance to understanding the
entire quasar phenomenon: (1) we need to understand how the
accretion/outflow processes work in AGNs and (2) we need to understand
the geometry and kinematics of the BLR to assess possible
systematic uncertainties in AGN black-hole mass measurements.

\section{Reverberation Mapping Basics}
Simply put, the idea behind reverberation mapping is to learn about
the structure and kinematics of the BLR by observing the detailed
response of the broad emission lines to changes in the continuum. The
basic assumptions needed are few and straightforward, and can largely
be justified after the fact:
\begin{enumerate}
\item {\it The continuum originates in a single central source.} The
size of the accretion disk in a typical bright Seyfert galaxy is
expected to be of order $10^{13}$-- $10^{14}$\,cm, or about a factor
of 100 or so smaller than the BLR turns out to be. It is worth noting that
we do not necessarily have to assume that the continuum is emitted
isotropically.
\item {\it Light-travel time $\tau_{\rm LT} = r/c$
is the most important time scale.} The other
potentially important time scales include:
\begin{itemize}
\item The recombination time scale $\tau_{\rm rec} = 
(n_e \alpha_{\rm B})^{-1}$, 
which is the time for emission-line gas to re-establish
photoionization equilibrium in response to a change in the continuum
brightness. For typical BLR densities, $n_e \approx 10^{10}\,{\rm
cm}^{-3}$, $\tau_{\rm rec} \approx 0.1$\,hr, i.e., virtually
instantaneous relative to the light-travel timescales of days to weeks
for luminous Seyfert galaxies.
\item The dynamical time scale for the BLR gas, $\tau_{\rm dyn}
\approx r/\Delta V$. For typical luminous Seyferts,
this works out to be of order 3--5 years. Reverberation experiments
must be kept short relative to the dynamical timescale to avoid
smearing the light travel-time effects.
\end{itemize}
\item {\it There is a simple, though not necessarily linear, relationship
between the observed continuum and the ionizing continuum.} In
particular, the observed continuum must vary in phase with the
ionizing continuum, which is what is driving the line variations. This
is probably the most fragile of these assumptions since
there is some evidence that long-wavelength continuum variations
follow those at shorter wavelengths, but the timescales involved are
still significantly shorter than the timescales for emission-line
response.
\end{enumerate}
 
Given these assumptions, a linearized response model can be written as
\begin{equation}
\Delta L(V,t) = \int \Psi(V,\tau) \Delta C(t-\tau)\,d\tau,
\end{equation}
where $\Delta C(t)$ is the continuum light curve relative to its mean
value $\overline{C}$, i.e., $\Delta C(t) = C(t) - \overline{C}$, and,
$\Delta L(V,t)$ is the emission-line light curve as a function of
line-of-sight Doppler velocity $V$ relative to its mean value
$\overline{L}(V)$. The
function $\Psi(V,\tau)$ is the ``velocity-delay map,'' i.e., the BLR
responsivity mapped into line-of-sight velocity/time-delay space. It
is also sometimes referred to as the ``transfer function" and eq.\ (1)
is called the transfer equation. Inspection of this formula shows that
the velocity-delay map is essentially the observed response to a
delta-function continuum outburst.  This makes it easy to construct
model velocity-delay maps from first principles.

Consider first what an observer at the central source would see 
in response to a
delta-function (instantaneous) outburst. Photons from the outburst
will travel out to some distance $r$ where they will be intercepted
and absorbed by BLR clouds and produce emission-line photons in
response. Some of the emission-line photons will travel back to the
central source, reaching it after a time delay $\tau =2r/c$. Thus a
spherical surface at distance $r$ defines an ``isodelay surface'' since
all emission-line photons produced on this surface are observed to
have the same time delay relative to the continuum outburst. 
For an observer at any
other location, the isodelay surface would be the locus of points for
which the travel from the common initial point (the continuum source) to
the observer is constant. It is obvious that such a
locus is an ellipsoid. When the observer is moved to infinity, the
isodelay surface becomes a paraboloid. We show a typical isodelay
surface for this geometry in the top panel of Figure 1.

\begin{figure}
\begin{center}
\includegraphics[height=10cm]{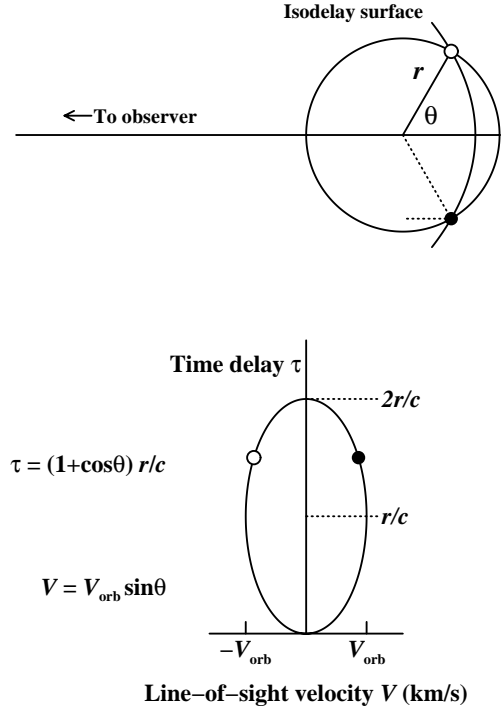}
\caption{{\em Upper diagram:} In this simple, illustrative model,
the line-emitting clouds are taken to be on a circular orbit of
radius $r$ around the central black hole. The observer is to the
left. In response to an instantaneous continuum outburst, the
clouds seen by the distant observer at a time delay $\tau$ after
detection of the continuum outburst will be those that lie
along an ``isodelay surface,'' for which the time delay relative
to the continuum signal will be $\tau = (1 + \cos \theta)r/c$,
the length of the dotted path shown. {\em Lower diagram:}
The circular orbit is mapped into the line-of-sight 
velocity/time-delay plane.}
\end{center}
\end{figure}

We can now construct a simple velocity delay map. Consider the trivial
case of BLR that is comprised of an edge-on (inclination 90$^{\rm o}$) ring
of clouds in a circular Keplerian orbit, as shown on the top panel of
Figure 1. In the lower panel of Figure 1, we map the points from polar
coordinates in configuration space to points in velocity-time delay
space. Points ($r$, $\theta$) in configuration space
map into line-of-sight velocity/time-delay space ($V$, $\tau$) according
to $V = -V_{\rm orb}\sin \theta$, where $V_{\rm orb}$ is
the orbital speed, and $\tau = (1 + \cos \theta)r/c$.
Inspection of Figure 1 shows that a circular Keplerian orbit
projects to an ellipse in velocity-time delay space. Generalization to
radially extended geometries is simple: a disk is a system of rings of
different radii and a spherical shell is a system of rings of
different inclinations. Figure 2 shows a system of circular Keplerian
orbits, i.e., $V_{\rm orb}(r) \propto r^{-1/2}$, and how these
project into velocity-delay space. A key feature of Keplerian systems
is the ``taper'' in the velocity-delay map with increasing time delay.

\begin{figure}
\begin{center}
\includegraphics[height=8cm]{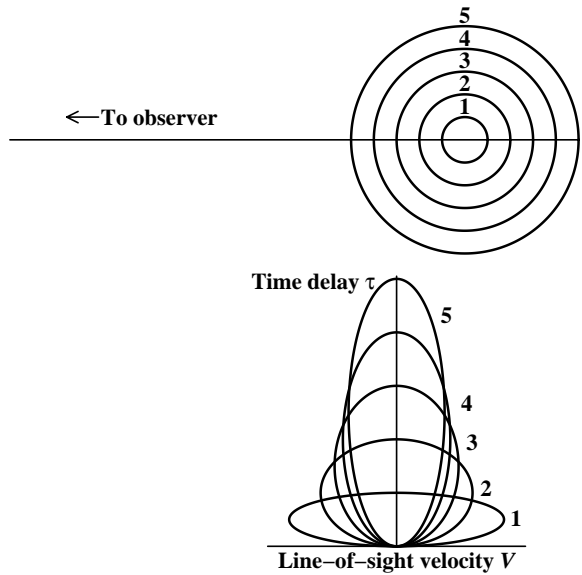}
\caption{This diagram is similar to Figure 1. Here we show
how circular Keplerian orbits of different radii map into
the velocity-time delay plane. Inner orbits have a larger
velocity range ($V\propto r^{-1/2}$) and shorter range
of time delay ($\tau_{\rm max} = 2r/c$), resulting in 
tapering of the map in velocity with increasing time
delay, a general feature of gravitationally dominated systems.}
\end{center}
\end{figure}

In Figure 3, we show two complete velocity-delay maps for radially
extended systems, in one case a Keplerian disk and in the other a
spherical system of clouds in circular Keplerian orbits of random
inclination. In both examples, the velocity-delay map is shown in the
upper left panel in greyscale. The lower left panel shows the result
of integrating the velocity-delay map over time delay, thus yielding
the emission-line profile for the system. The upper right panel shows
the result of integrating over velocity, yielding the total time
response of the line; this is referred to as the ``delay map'' or
the ``one-dimensional transfer function.'' Inspection of Figure
3 shows that these two velocity-delay maps are superficially similar;
both show clearly the tapering with time delay that is characteristic
of Keplerian systems and have double-peaked line profiles.
However, it is also clear that they can be
easily distinguished from one another. This, of course, is the key:
the goal of reverberation mapping is to use the observables, namely
the continuum light curve $C(t)$ and the emission-line light curve
$L(V,t)$ and invert eq.\ (2) to recover the velocity-delay map
$\Psi(V,\tau)$. Equation (2) represents a fairly common type of problem
that arises in many applications in physics and engineering. Indeed,
the velocity-delay map is the Green's function for the
system. Solution of eq.\ (2) by Fourier transforms immediately
suggests itself, but real reverberation data are far too sparsely
sampled and usually too noisy to this method to be effective. Other
methods have to be employed, such as reconstruction by the maximum
entropy method (Horne 1994).  Unfortunately, even the best
reverberation data obtained to date have not been up to the task
of yielding a high-fidelity velocity-delay map. Existing
velocity-delay maps are noisy and ambiguous. Figure 4 
shows the result of an attempt to recover a 
velocity-delay map for the C\,{\sc iv}--He\,{\sc ii}\
spectral region in NGC 4151 (Ulrich \& Horne 1996). The Keplerian
taper of the map is seen, but other possible structure is only hinted
at, as it is in other attempt to recover a velocity-delay map from
real data (e.g., Wanders et al.\ 1995; Done \& Krolik 1996;
Kollatschny 2003).  It must be pointed out, however, that no case to
date has recovery of the velocity-delay map been a design goal for an
experiment. Previous reverberation-mapping experiments have had the
more modest goal of recovering only the mean response time of emission
lines, from which one can still draw considerable information.  By
integrating eq.\ (2) over velocity and then convolving it with the
continuum light curve, we find that under reasonable conditions,
cross-correlation of the continuum and emission-line light curves
yields the mean response time, or ``lag,'' for the emission lines.

\begin{figure}
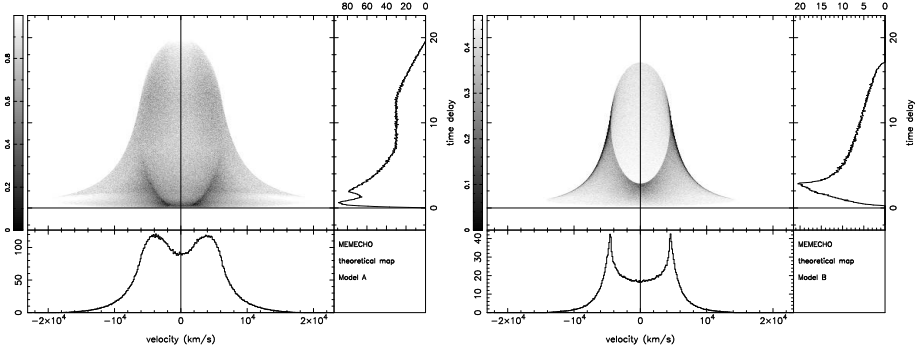

\begin{center}
\includegraphics[height=6cm,angle=-90]{figure3a.ps}
\includegraphics[height=6cm,angle=-90]{figure3b.ps}
\caption{Theoretical velocity-delay maps $\Psi(V,\tau)$ 
shown in greyscale for
a spherical distribution of line-emitting clouds
in circular Keplerian orbits of random inclination (left)
and an inclined Keplerian disk of line-emitting clouds (right).
Projections in velocity and time-delay show the 
line profile (below) and delay map (right).
From Horne et al.\ (2004).}
\end{center}
\end{figure}

\begin{figure}
\begin{center}
\includegraphics[height=8cm,angle=-90]{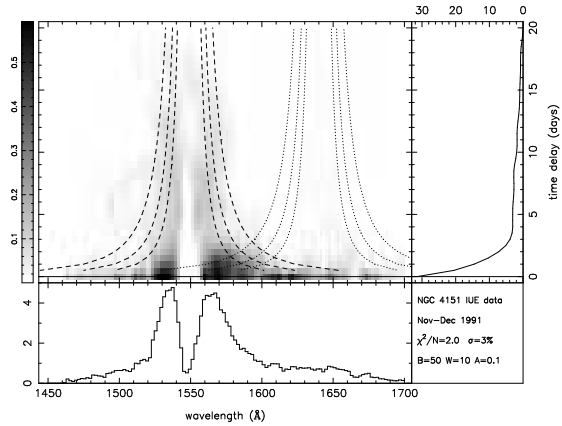}
\caption{A velocity--delay map for the C\,{\sc iv}--He\,{\sc ii}
region in NGC 4151, based on data obtained with the 
{\it International Ultraviolet Explorer}. From
Ulrich \& Horne (1996).}
\end{center}
\end{figure}

\section{Reverberation Results} 
Prior to about 1988, there were a large number of observations that
suggested that the broad emission lines in Seyferts varied in response
to continuum variations and did so on surprisingly short time
scales. These early results led to the first highly successful 
reverberation campaign, carried
out in 1988--89, combining UV spectra obtained with the {\it
International Ultraviolet Explorer (IUE)} with ground-based optical
observations from numerous observatories. The program ran for over 8
months and achieved time resolution of a few days in several continuum
and emission-line time series (Clavel et al. 1991; Peterson et al.\
1991; Dietrich et al.\ 1993; Maoz et al.\ 1993). A number of important
results were produced by this project, including:
\begin{enumerate}
\item From the shortest measured wavelength (1350\,\AA) to the longest
(5100\,\AA), the continuum variations appear to be in phase, with any
lags between bands amounting to no more than a couple of days.
\item The highest ionization emission lines respond most rapidly to
continuum variations (e.g., $\sim 2$\,days for He\,{\sc ii}\,$\,\lambda1640$
and $\sim10$\,days for Ly$\alpha$\ and 
C\,{\sc iv}\,$\lambda1549$) and the lower
ionization lines respond less rapidly (e.g., $\sim20$\,days for
H$\beta$\ and nearly 30 days for C\,{\sc iii}]\,$\lambda1909$).  
The BLR thus shows radial ionization stratification.
\end{enumerate}

Optical spectroscopic monitoring of NGC\,5548 continued for a total of
13 years, and during the fifth year of the program (1993), concurrent
high-time resolution (daily observations) were made for about 60 days
with {\em IUE} and for 39 days with the Faint Object Spectrograph on
{\em Hubble Space Telescope} (Korista et al.\ 1995). Over time, it
became clear that the H$\beta$\ emission-line lag is a {\it dynamic}
quantity, it varies with time and is dependent on the current mean
continuum luminosity (Peterson et al.\ 2002).
In other words, there is much more nuclear gas on scales of
thousands of gravitational radii than previously thought: at any given
time, most of the emission in any particular line arises primarily in
that gas for which the physical conditions optimally produce that
particular emission line (cf.\ the ``locally optimized cloud'' model of
Baldwin et al.\ 1995).

Peterson et al.\ (2004) recently completed a comprehensive 
reanalysis of 117 independent reverberation mapping data sets 
on 37 AGNs, measuring emission-line lags, line widths,
and black hole masses for all but two of these sources.
Calibration of the reverberation-based
mass scale, as embodied in the scaling factor $f$ in eq.\ (1), is set
by assuming that AGNs follow the same relationship between black hole
mass and the host-galaxy bulge velocity dispersion (the $M_{\rm
BH}-\sigma_*$ relationship) seen in quiescent galaxies (Onken et al.\
2004). The range of measured masses runs from 
$\sim2\times10^6\,M_{\odot}$ for the
narrow-line Seyfert 1 galaxy NGC 4051 to $\sim1.3\times10^9\,M_{\odot}$
for the quasar PG 1426+015. 
The statistical errors in the mass measurements (due
to uncertainties in lag and line-width measurement) are typically only about
30\%. However, the systematic errors, due to scatter in the 
$M_{\rm BH}-\sigma_*$ relationship, amount to about a factor of three; this
systematic uncertainty can decreased only by understanding the geometry and
kinematics of the BLR. Figure 5 shows a current version of the
mass-luminosity relationship for AGNs, based on these
reverberation-based black hole masses.

\begin{figure}
\begin{center}
\includegraphics[height=7cm]{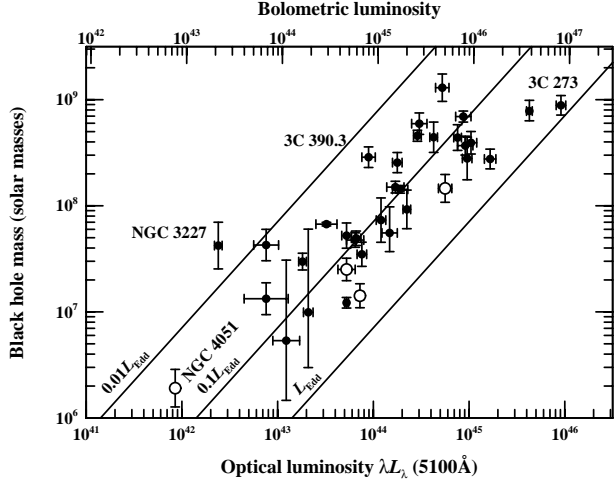}
\caption{Black hole mass vs.\ luminosity for 35 
reverberation-mapped AGNs. The luminosity scale on the lower x-axis is
$\log \lambda L_{\lambda}$ in units of ergs s$^{-1}$. The
upper x-axis shows the bolometric luminosity assuming that
$L_{\rm bol} \approx 9 \lambda L_{\lambda}$. The diagonal
lines show the Eddington limit $L_{\rm Edd}$,
$0.1 L_{\rm Edd}$, and $0.01L_{\rm Edd}$. The open circles
represent narrow-line Seyfert 1 galaxies. From Peterson
et al.\ (2004).}
\end{center}
\end{figure}

\section{The Future: What Will It Take to Map the Broad-Line Region?}
While we still do not have a velocity-delay map in hand, we certainly
know how to get one. More than a dozen years of experience in
reverberation mapping have led to a reasonably good understanding of
the timescales for response of various lines as a function of
luminosity and of how the continuum itself varies with time. On the
basis of this information, we have carried out extensive simulations
to determine the observational requirements to obtain high-fidelity
velocity-delay maps for emission lines in moderate luminosity Seyfert
galaxies (quantities that follow are based specifically on NGC 5548,
by far the AGN best studied by reverberation). A sample numerical
simulation is shown in Figure 6. As described more
completely by Horne et al.\ (2004), the principal requirements are:
\begin{enumerate}
\item {\it High time resolution,} (less than 0.2--1\,day, depending
on the emission line).  The interval between observations translates
directly into the resolution in the time-delay axis.
\item {\it Long duration (several months).} A rule of thumb in time
series analysis is that the duration of the experiment should exceed
the maximum timescale to be probed by at least a factor of three. The
lag for H$\beta$\ in NGC 5548 is typically around 20 days, so the
longest timescale to be probed is $2r/c$. The duration should thus be
at least $\sim120$\,days to map the H$\beta$-emitting region.
However, since C\,{\sc iv} seems to respond twice as fast as
H$\beta$, the C\,{\sc iv}-emitting region might be mapped in 
as little as $\sim 60$ days. A more important consideration,
however, is detection in the time 
series a strong continuum signal, such as a change in
sign of the derivative of the light curve. This produces a similarly
strong emission-line response. We find that $\sim200$\,days
of observations are required {\it to be certain} that such an event occurs, and
to observe its consequences in the emission lines.
\item {\it Moderate spectral resolution} ($\leq 600$\,km\,s$^{-1}$). While
higher spectral resolution is always desirable, AGN emission lines
show little additional structure at resolution better than several hundred
kilometers per second. Higher resolution does, however, make it in
principle possible to detect a gravitational redshift 
(e.g., Kollatschy 2004), providing an independent and complementary 
measure of the black hole mass.
\item {\it High homogeneity and signal-to-noise ratios} 
($S/N \approx 100$). Both
continuum and emission-line flux variations are small on short time
scales, typically no more than a few percent on diurnal
timescales. Excellent {\it relative} flux calibration and
signal-to-noise ratios are necessary to make use of the high time
resolution.
\end{enumerate}
\begin{figure}
\begin{center}
\includegraphics[height=16cm]{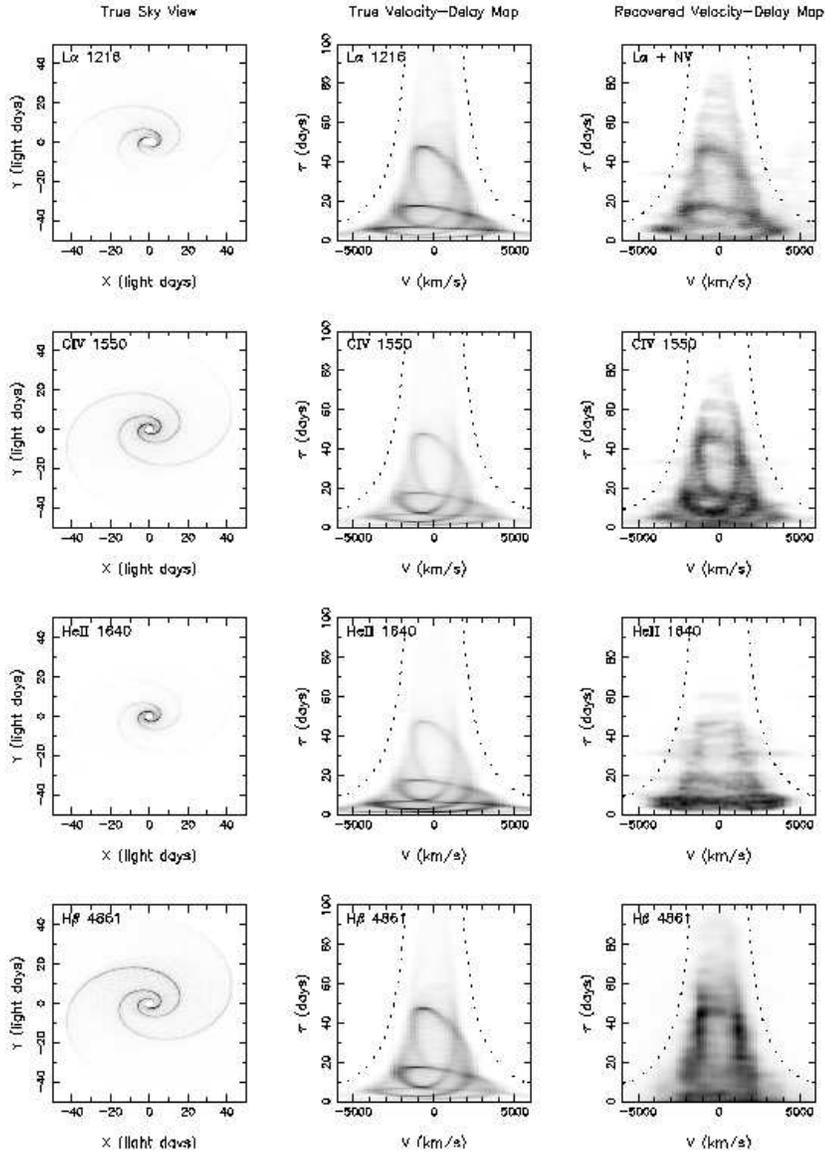}
\caption{Numerical simulations of velocity--delay map
recovery. An arbitrary but complex geometry was chosen
to show that a complicated BLR structure can be recovered.
The left column shows the BLR geometry in four
lines, Ly$\alpha\,\lambda1215$, C\,{\sc iv}\,$\lambda1549$,
He\,{\sc ii}\,$\lambda1640$, and H$\beta\,\lambda4861$, from
top to bottom. The middle column shows the BLR model mapped
into velocity--delay space. The right column shows the
recovered velocity--delay map recovered from
simulated data. From
Horne et al.\ (2004).}
\end{center}
\end{figure}

The final point makes it clear that this will be difficult to do with
ground-based observations: it is hard to maintain such high levels of
homogeneity and thus accurate relative flux calibration with a
variable point-spread function, such as one dominated by the effects
of atmospheric seeing. Space-based observations are much more likely
to succeed, and moreover, the ultraviolet part of the spectrum gives
access to the important strong lines Ly$\alpha$,
N\,{\sc v}\,$\lambda1240$, Si\,{\sc iv}\,$\lambda1400$, 
C\,{\sc iv}\,$\lambda1549$, and
He\,{\sc ii}\,$\lambda1640$, all of which vary strongly.

We have carried out a series of simulations assuming specifically
observation of NGC 5548 with the Space Telescope Imaging Spectrograph
(STIS) on {\em HST}. We assume that the typical BLR response times are
as observed during the first major monitoring campaign in 1989, as
these values appear to be typical. We also assume for practical
reasons  that we can obtain only one observation per day
(mostly on account of restrictions against the use of the STIS
UV detectors during orbits when {\em HST} passes through
the South Atlantic Anomaly) which we must complete in one {\em HST}
orbit, using the typical time that NGC 5548 is observable per
orbit. Furthermore, we included in the simulations nominal spacecraft
and instrument safing events of typical duration (generally a few
days) and frequency. 
We also considered the effects of early termination of the
experiment due, for example, to a more catastrophic failure. We
carried out 10 individual simulations, with each one using a different
continuum model; all of the continuum models were "conservative" in
the sense that the continuum activity is weaker and less pronounced
than it is usually observed to be.

Given our assumptions, we find that all 10 of the simulated
experiments succeed within 200 days. For the most favorable continuum
behavior, success can be achieved in as little as $\sim60$\,days,
though this is rare, or more commonly in around $\sim150$\, days.  We
also find that these results are robust against occasional random data
losses.

These generally conservative assumptions argue strongly that
velocity-delay maps for all the strong UV lines in NGC 5548 could be
obtained in a 200-orbit {\it HST} program based on one orbit per
day. There are clearly elements of risk associated with the program,
the most obvious being early termination on account of a systems
failure or a major safing event that would end the time series
prematurely. This risk is somewhat mitigated by the conservatism of
our simulations; it is possible that the experiment could succeed in
much less time. Indeed, if we define ``success'' as obtaining a
velocity-delay map that is stable for 50 days, we find that the
probability of success is as high as $\sim90$\% in 150 days.

Finally, one might also ask about the scientific risk: for example,
what if the velocity-delay map, though of high fidelity, cannot be
interpreted? In other words, what if the velocity-delay map is a
``mess'' and has no discernible structure? First of all, this is not a
likely outcome since long-term monitoring shows persistent features in
emission-line profiles that imply there is some order or symmetry to
the BLR. Moreover, even if the BLR turns out to be a ``mess'' we'll
still have learned an important fact about the BLR structure, namely
that it is basically chaotic. But the
bottom line is that right now we have {\it nearly complete ignorance}
about the BLR structure. We cannot even assess to {\it any} level of
confidence how many velocity-delay maps of AGNs we will need to solve
the problem until we have obtained at least {\it one} velocity-delay
map of at least {\it one} emission line. Until we have that, our
knowledge of the role of the BLR in AGN fueling and outflows will
remain based on theoretical speculation alone, historically a very
dangerous situation for astrophysicists.

\section{Summary} 
We have argued that reverberation mapping provides a unique probe of
the inner structure of AGNs. The reverberation technique has been very
successful in determining the BLR sizes and black hole masses in 35
AGNs. The masses obtained are accurate to about a factor of 3,
based on the observed scatter in the AGN $M_{\rm BH}-\sigma_*$
relationship.  The accuracy of these masses is fundamentally limited
by unknown geometry and kinematics of BLR. We have also argued that it
is possible to obtain complete, high-fidelity velocity-delay maps of
the strong ultraviolet lines in relatively nearby, moderately
luminous AGNs with {\it HST}, and we specifically argue that this can
be done with high confidence of success for NGC 5548
with one {\it HST} orbit per day for a period of no longer than 200 days.

\begin{acknowledgments}
We are grateful for support by the National Science Foundation
(grant AST--0205964) and PPARC.
\end{acknowledgments}

\end{document}